\documentclass[%
 reprint,
 amsmath,amssymb,
 aps,
]{revtex4-1}

\usepackage{graphicx}% Include figure files
\usepackage{dcolumn}% Align table columns on decimal point
\usepackage{bm}% bold math
\begin{document}

%\preprint{APS/123-QED}

\title{Surface and 3D quantum Hall effects from engineering of exceptional points in nodal-line semimetals}% Force line breaks with \\

\author{Rafael A. Molina}
%\email{rafael.molina@csic.es}
\author{Jos\'e Gonz\'alez}%
%\email{j.gonzalez@csic.es}
\affiliation{%
 Instituto de Estructura de la Materia - CSIC, Serrano 123, E-28006 Madrid, Spain}%

\date{\today}% It is always \today, today,

\begin{abstract}
We investigate the potential of the surface states of 3D nodal-line semimetals to produce surface and 3D quantized Hall effects in which the electronic currents flow unidirectionally along the surface of the material. This phenomenology is based on the fact that, irrespective of its orientation, a strong magnetic field gives rise to a zero Landau level with huge 2D degeneracy of states in the nodal-line semimetal.
When the magnetic field is perpendicular to the plane of the nodal line, we show that the evanescent states within the nodal ring are transmuted into Landau surface states which correspond to exceptional points, i.e. branch points which arise upon extension of the spectrum to complex values of the momentum. These Landau surface states are then topologically protected, as the branch cuts in the complex plane lead to a spectrum with the topology of a torus which cannot be modified by smooth perturbations. When the magnetic field is parallel to the nodal ring, we find instead that the evanescent waves are paired {\em inside} the bulk of the semimetal, leading to a 3D quantum Hall effect in which Landau states extended over 2D slices of the 3D bulk form a flat level, but start to get some dispersion as they approach the surface of the material. We show that, for both orientations of the magnetic field, the transverse Hall conductivity is quantized in units of $e^2/h$, with very large values of the Chern number which afford very high conductivities along the surface of the 3D material. 
\end{abstract}

\pacs{Valid PACS appear here}% PACS, the Physics and Astronomy
\maketitle

%\tableofcontents

{\em Introduction.---}
Currently, there is a huge interest in the new family of 3D topological semimetals which include Dirac and Weyl semimetals with isolated Dirac or Weyl nodes in the band structure \cite{Liu14,Neupane14,Borisenko14,Xu15}, and the nodal-line semimetals with a continuous line of nodes in the Brillouin zone \cite{Burkov16,Fang16}. There are already several experimentally established examples of topological nodal-line semimetals including PbTaSe$_2$ \cite{Bian16}, PtSn$_4$ \cite{Wu16}, and ZrSiS \cite{Schoop16}. Apart from the theoretical interest, as their low-energy excitations behave as relativistic fermions, they exhibit very interesting features like the presence of a 2D manifold of surface states forming nearly flat or very narrow bands, with the potential for the development of very strong correlation\cite{Jia16}.  

% Part of the introduction: Paragraph about Hall effect? 

Another relevant instance leading to a large degeneracy of electronic states arises in low-temperature 2D samples in the presence of strong magnetic field, where the Hall conductivity can be quantized as first discovered by Von Klitzing {\em et al.} \cite{Klitzing80}. 
%The impact of this quantum Hall effect in applications, specially in metrology was immediately realized. 
The explanation of this phenomenon is the paradigmatic example of application of topological concepts in condensed matter physics. The integer values dictating the quantization of the Hall conductivity can be written in terms of topological invariants known as Chern numbers, closely related to Berry phases \cite{Thouless82}. 
%Hall physics has been explored also in Dirac bidimensional materials, like graphene \cite{Novoselov07}. 

As long as the 2D quantum Hall effect is supported by boundary states, it turns out quite interesting to ask about the new physical effects which may arise from the plethora of surface states of a 3D nodal-line semimetal in a strong magnetic field. In this setting, it becomes relevant to investigate whether the manifold of surface states may build a quantized 3D Hall effect, and the kind of boundary states which may support the electronic transport in the 3D material.

Several studies have investigated the effect of a strong magnetic field in 3D semimetals \cite{Yang11,Cao15}. It has been proposed for instance that closed orbits can be formed in the bulk by connecting the Fermi arcs at opposite surfaces of 3D Weyl semimetals \cite{Potter14}. In the case of the nodal-line semimetals, it has been shown the possibility of having a zero-energy level and Landau bands following a pattern similar to that of Dirac fermions \cite{Rhim15}.

\begin{figure}[!hbt]
\includegraphics[width=7cm]{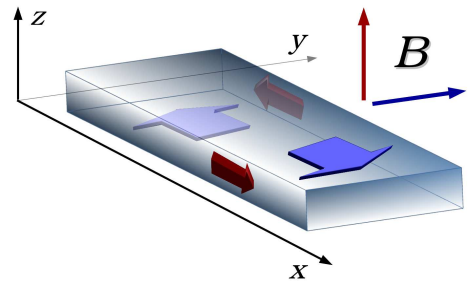}
\caption{Schematic plot of the surface electronic currents in a slab of 3D nodal-line semimetal for two orthogonal directions of the magnetic field.
% {\em Bottom panel: Schematic plot of the direction of the currents in the Landau levels a 3D slab of a nodal line semimetal depending on the direction of the magnetic field.}
}
\label{slab}
\end{figure}

In this paper, we unveil the potential of the 3D nodal-line semimetals to support a phenomenon which is the anologue of the 2D quantum Hall effect but promoted to a higher spatial dimension. We are going to see that, for different orientations of the magnetic field, there are 3D slab geometries where all the low-energy bulk states of the semimetal are localized while only the surface states contribute to the electronic transport. We can interpret this phenomenon as the conversion of the 3D nodal-line semimetal into a kind of topological insulator, with an orientation of the electronic currents along the surface which is dictated by the direction of the magnetic field (as shown schematically in Fig. \ref{slab}).

The reason for this conversion into a 3D quantum Hall system lies in the microscopic features of the surface states in the 3D nodal-line semimetals. Such states are evanescent waves with a pseudospin degree of freedom which can take opposite values at opposite faces of a slab geometry. This pseudospin is preserved when the magnetic field is perpendicular to the nodal ring, in such a way that the evanescent waves remain stabilized in a zero-energy level with huge 2D degeneracy (in momentum space) arising from the collapse down to zero-energy of a large number of 2D-like Landau levels. The evanescent waves of the semimetal are then transmuted into Landau surface states, which may get dispersion due to effects of lateral confinement (as in the slab geometry in Fig. \ref{slab}).

\begin{figure}[!hbt]
\includegraphics[width=6cm]{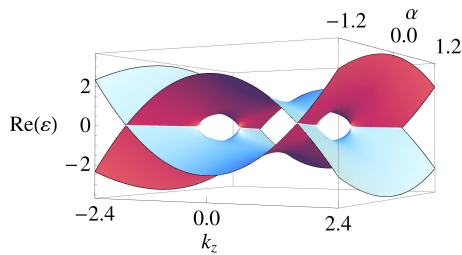}
\caption{Spectrum of a nodal-line semimetal for complex momenta $k_z + i\alpha$ ($k_z$ being perpendicular to the nodal ring) showing two branch cuts which connect exceptional points corresponding to evanescent zero-energy modes with opposite decay ($\pm \alpha $) along the direction $k_z$.
% {\em Bottom panel: Schematic plot of the direction of the currents in the Landau levels a 3D slab of a nodal line semimetal depending on the direction of the magnetic field.}
}
\label{bcuts}
\end{figure}

We will see that there is a deep reason for the topological protection of the Landau surface states in the zero-energy level of the nodal-line semimetals. Such states correspond to exceptional points, i.e. branch points which arise upon extension of the spectrum to complex values of the momentum \cite{Berry04,Heiss12}. The Landau surface states are then topologically protected, as the branch cuts in the complex plane lead to a spectrum with the topology of a torus (as represented in Fig. \ref{bcuts}) which cannot be modified by smooth perturbations. It is useful to remark that the Fermi arcs in Weyl semimetals and the drumhead states of nodal-line semimetals can also be described through exceptional points \cite{Gonzalez17}, as well as other rotating surface states appearing in Dirac and Weyl semimetals upon illumination by circularly polarized light \cite{Gonzalez16}.

Quite remarkably, we find that a magnetic field parallel to the nodal ring has the effect of pairing the evanescent waves {\em inside} the bulk of the semimetal, leading to states that reside in quasi-2D slices parallel to the magnetic field. This constitutes then a perfect 3D version of a quantum Hall effect, in which the states with such a 2D extension form a flat level in the bulk but start to get some dispersion as they approach the surface of the semimetal. This explains that the electronic transport takes place through surface states which are the analogue of the edge excitations for this kind of 3D Hall effect (as shown in the scheme of Fig. \ref{slab}). We will see that in this case, as well as when the magnetic field is perpendicular to the nodal ring, the transverse Hall conductivity is quantized in units of $e^2/h$ as $\sigma_{ij} = N e^2/h$, with very large values of the Chern number $N$ which afford very high conductivity along the surfaces of the nodal-line semimetal.

{\em Surface Hall effect.---}
Our starting point is a continuum model of nodal-line semimetal, whose suitability is well documented from the description of the compounds of the CaP$_3$ family \cite{Xu16}. In a system of units in which $\hbar=1$ (we recover fully dimensional units when discussing conductance in terms of the conductance quantum $e^2/h$) the Hamiltonian is given by
\begin{equation}
H_{\rm NL} = (m_0 + m_1 \boldsymbol{\nabla}^2 ) \sigma_z - iv \partial_z \sigma_x
\label{ham0}
\end{equation}
where $\sigma_i$ ($i=x,y,z$) stand for the Pauli matrices. In terms of the 3D momentum $\mathbf{k}$, the model displays two bands with energy
\begin{equation}
\varepsilon = \pm \sqrt{(m_0 - m_1 \mathbf{k}^2)^2 + v^2 k_z^2 }
\label{eq:energy}
\end{equation}
We find a line of nodes where the two bands meet in the plane $k_z = 0$, corresponding to the circular set $k_x^2 + k_y^2 = m_0 /m_1 $. A remarkable feature is that the projection of the nodal ring onto a given surface leads to a so-called drumhead, filled by surface states forming in general nearly flat or very narrow bands.

In the presence of a magnetic field in the $z$ direction (perpendicular to the plane of the nodal line), the vector potential can be written as
$\mathbf{A} = (-By,0,0)$. Setting units in which $e = 1$ and $c = 1$, the Hamiltonian of the nodal line semimetal becomes
\begin{equation}
H_{\perp }  = \left[m_0 + m_1 \left( - (-i\partial_x - By)^2 + \partial_y^2 + \partial_z^2 \right) \right] \sigma_z  
    -iv \partial_z  \sigma_x 
\label{H1}
\end{equation}
The eigenstates in the bulk of the semimetal are given in terms of the eigenfunctions of a harmonic oscillator centered at $y = k_x/B$, with energy eigenvalues
\begin{equation}
\varepsilon_n = \pm \sqrt{[m_0 - m_1 k_z^2 - 2 m_1 B (n+1/2)]^2 + v^2 k_z^2},
\end{equation}
where $n \geq 0$ is the Landau level index.

It is interesting the search of evanescent states decaying from a 2D boundary of the semimetal (at $z = 0$) as
\begin{equation}
\Psi  \sim  e^{ik_z z} e^{-\alpha z} \chi (x,y)
\end{equation}
This converts the eigenvalue problem into the equation
\begin{eqnarray}
\left[ m_0 - m_1\left( 2B(n+1/2) + k_z^2 - \alpha^2 + 2ik_z\alpha \right) \right] \sigma_z \chi \nonumber \\
 + v(k_z + i\alpha) \sigma_x \chi = \varepsilon \chi
\label{ev}
\end{eqnarray}
A zero-energy eigenvalue can be obtained by taking $\chi $ such that $\sigma_y \chi = \chi $. Then, in the model with $4m_0 m_1 > v^2$, Eq. (\ref{ev}) admits a solution with $\alpha = v/2m_1 $ and values of $k_z$ given by 
\begin{equation}
k_z = \sqrt{\frac{m_0}{m_1} - 2B(n+1/2) - \alpha^2 }
\label{eva1}
\end{equation}
Otherwise, in the model with $4m_0 m_1 < v^2$, the zero-energy evanescent states correspond to $k_z = 0$ and values of $\alpha $ given by 
\begin{equation}
\alpha=\frac{v \pm \sqrt{v^2 - 4m_1(m_0 - m_1 2B(n+1/2))}}{2m_1}
\label{eva2}
\end{equation}

This construction reveals that, in the presence of a magnetic field perpendicular to the plane of the nodes, the drumhead surface states are converted into Landau states decaying from the surface of the nodal-line semimetal. In general, this transmutation leads to the collapse of a great number of Landau levels down to zero energy. This degeneracy applies to Landau levels with order $n$ such that $m_0 - 2 m_1 B (n+1/2) > 0$, up to a maximum value beyond which the conditions of evanescence (\ref{eva1}) and (\ref{eva2}) cannot be fulfilled.

\begin{figure}[!hbt]
\includegraphics[width=7cm]{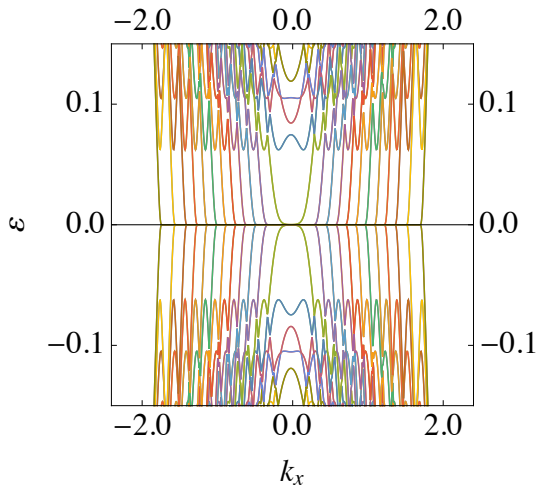}
\caption{Band structure for a slab of nodal-line semimetal of width $W=40$ nm (lateral dimension $\Delta y = 40$ nm), for parameters $m_0=1.2$ eV, $m_1=1.0$ eV nm$^2$, $v=0.5$ eV nm, and magnetic field (perpendicular to the nodal ring) $B = 30$ T. The energy is measured in eV and the momentum in nm$^{-1}$.}
\label{bands}
\end{figure}

The band structure for a slab of width $W = 40$ nm (in the $z$ direction) in a magnetic field with $B = 30$ T is represented in Fig. \ref{bands}. The plot illustrates the huge degeneracy of Landau states in the zero-energy level, which may include in general the superposition of a large number of Landau levels with different order $n$.

An important feature of the surface Landau states is that they are topologically protected by the complex structure developed upon extension of the momentum to complex values $k_z + i\alpha $. The surface Landau states correspond to exceptional points in the spectrum of the Hamiltonian, which are identified as square-root singularities in the complex plane $(k_z, \alpha )$. This can be seen in Fig. \ref{bcuts}, which shows the Riemann sheet of the spectrum of a nodal-line semimetal for $n=0$. The complex structure has two branch cuts endowing the spectrum with the topology of a torus, which cannot be altered unless the branch cuts are closed by pair-wise annihilation of the exceptional points.

The total number of exceptional points can be indeed controlled with the strength of the magnetic field $B$. In the model with $4m_0 m_1 > v^2$ (that corresponds to a type A nodal-line semimetal according to the classification put forward in Ref. \cite{Gonzalez17}), there is a series of branch cuts across the real axis $k_z$, of the same type as those shown in Fig. \ref{bcuts}. These branch cuts can be moved towards the imaginary axis by increasing the value of $B$. When a couple of exceptional points with opposite values of $\alpha $ reach that axis, they continue moving towards the origin, until they coalesce and mutually annihilate. This process can be repeated for all the branch cuts, leading to the disappearance of all the exceptional points (and the associated surface Landau states) at a sufficiently large magnetic field with $B>m_0/m_1$. For a type B nodal-line semimetal with $4m_0m_1<v^2$, there is a similar process, although the branch cuts start running along the imaginary axis, and the Landau states disappear again for $B>m_0/m_1$. 

The surface Landau states may support a quantum Hall effect in the 3D semimetal, with surface currents along the vertical boundary of a slab with finite horizontal dimension $\Delta y$ as shown in Fig. \ref{slab}. It is worthwhile to note that the evanescent solutions obtained from Eq. (\ref{ev}) (valid in the limit of infinite depth $W$ along the $z$ direction) cannot carry a current along the $x$ direction, as the current operator is given by $j_x = -2m_1 (k_x - By) \sigma_z $ while the evanescent waves are eigenstates of $\sigma_y $. However, this does not hold in the case of a slab with finite width $W$, as the two evanescent waves attached to opposite faces of the slab start to hybridize when approaching the boundaries of the lateral dimension $\Delta y$. This explains the dispersion of the bands from the zero-energy Landau level in Fig. \ref{bands}, which implies a nonvanishing current as $\langle j_x \rangle = \partial \varepsilon / \partial k_x $. The $N$-fold degeneracy of the zero-energy level leads then to a transverse Hall conductivity $\sigma_{xy} = N (e^2/h)$, where the number of channels may be as large as $N \sim 30$ for $B = 10$ T and parameters $m_0 \sim 1$ eV, $m_1 \sim 1.0$ eV nm$^2$.

{\em 3D Hall effect.---}
We consider now the case in which the magnetic field is parallel to the plane of the nodal line, taking for definiteness a constant field pointing in the $y$ direction. The vector potential can be chosen as $\mathbf{A} = (Bz,0,0)$ and the Hamiltonian reads in that gauge
\begin{equation}
H_{\parallel }  =  \left[m_0 + m_1 \left( - (-i\partial_x+Bz)^2 + \partial_y^2 + \partial_z^2 \right) \right] \sigma_z  
    -iv \partial_z  \sigma_x 
\end{equation}

\begin{figure}[!hbt]
\includegraphics[width=5cm]{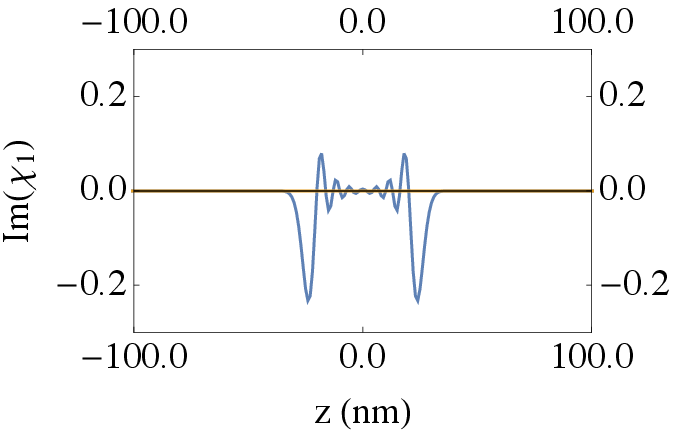}
\includegraphics[width=5cm]{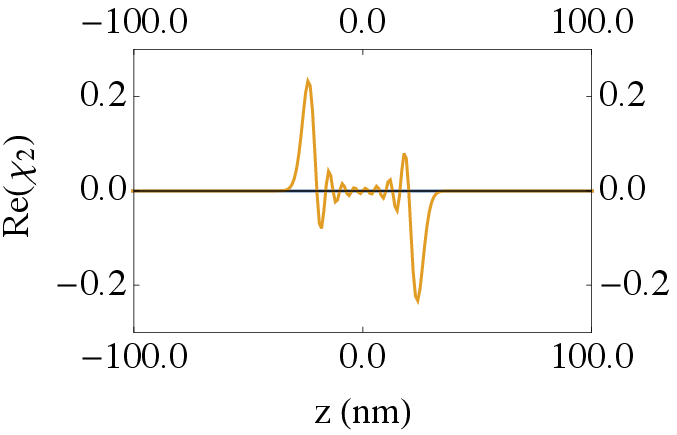}
\caption{Plot of the wave function along the direction perpendicular to the nodal ring (zoom view) showing (a) the imaginary part of the first component (as ${\rm Re}(\chi_1) = 0$), and (b) the real part of the second component (as ${\rm Im}(\chi_2) = 0$) of a state with $k_x = k_y = 0$ in the zero Landau level of a nodal-line semimetal with $m_0=2.0$ eV, $m_1=1.5$ eV nm$^2$, $v=0.5$ eV nm, and magnetic field (parallel to the nodal ring) $B = 30$ T. }
\label{wavef}
\end{figure}

In this case, the most interesting effects manifest again in a slab with finite width $W$ in the $z$ direction. In this geometry, there is a huge set of eigenstates forming a Landau level with virtually zero energy. Quite remarkably, those states are localized at different 2D slices within the bulk of the slab. A typical shape of wave function along the $z$ direction for a state in the zero Landau level can be seen in Fig. \ref{wavef}. The localization of the state in the bulk can be moved by varying the momentum $k_x$, which shifts the position of the slice within the slab.

The unconventional character of the quasi-2D states in the bulk is remarked by the fact that they arise from the superposition of two waves, both evanescent along $z$ but with opposite orientations. The possibility of having those waves can be realized from inspection of the eigenvalue problem (for $k_x = k_y = 0$)
\begin{equation}
\left[m_0 + m_1 \left( \partial_z^2 - B^2 z^2  \right) \right] \sigma_z \chi  
    -iv \partial_z  \sigma_x  \chi  =  \varepsilon \chi
\label{ep}
\end{equation}
Eq. (\ref{ep}) looks like a Dirac equation for massive fermions in which the magnetic field provides a confining potential. 
Thus, the potential well gives rise to two domain walls (turning points) along the $z$ direction which are able to pin the evanescent waves, leading to localized eigenstates within the gap opened by the mass in the Dirac spectrum.

\begin{figure}[!hbt]

%\includegraphics[width=8cm]{NLtypeA20t.png}
%\begin{center}
%(a)
%\end{center}
\includegraphics[width=7cm]{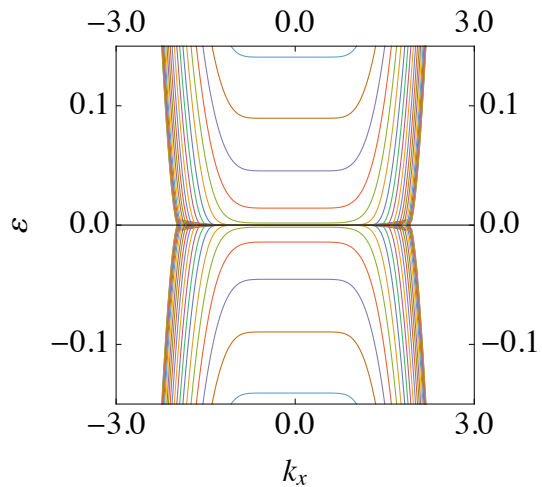}
\begin{center}
%(b)
\end{center}
%\caption{Band structure of nodal-line semimetal (a) for a slab of width $W=120$ nm, parameters $m_0=0.5$ eV,
%$m_1=1$ eV nm$^2$, $v=1$ eV nm, and magnetic field $B=20$ T, (b) for a slab of width $W=40$ nm (lateral dimension $\Delta y = 40$ nm), parameters $m_0=0.64$ eV, $m_1=1$ eV nm$^2$, $v=1$ eV nm, and magnetic field $B \approx 41$ T.}
\caption{Band structure for a slab of nodal-line semimetal of width $W=60$ nm (lateral dimension $\Delta y = 60$ nm), for parameters $m_0=0.5$ eV, $m_1=0.5$ eV nm$^2$, $v=0.5$ eV nm, and magnetic field (parallel to the nodal ring) $B = 30$ T. The units are the same as in Fig. \ref{bands}.}
\label{bslab}
\end{figure}

It can be checked that the highly degenerate level at virtually zero energy arises from the collapse of a large number of flat bands corresponding to different values of the momentum $k_y$. This is represented in Fig. \ref{bslab}, where we can see the band structure for a slab of width $W = 60$ nm in the $z$ direction, under a magnetic field with $B = 30$ T. 
%Tenemos que cambiar esto para las nuevas figuras. 
The momentum $k_x$ is used in the plot to label the states in each of the bands, which ultimately start to disperse when the quasi-2D slices supporting the states approach the faces of the slab.

We can make contact at this point with the correspondence between evanescent waves and exceptional points of the previous section. In this respect, the pairing of the two evanescent waves in each quasi-2D slice can be viewed as a binding of exceptional points with opposite values $\pm \alpha $ of the imaginary part of the momentum. In this case, the evanescent waves in each pair are subjected to interaction, which can be amplified modifying the values of $k_x$ and $k_y$, bringing for instance the pair of evanescent waves closer to the faces of the slab. Indeed, the energy of a localized state in the bulk is never exactly zero (though being very small) due to the binding of the two evanescent waves. Increasing the field strength $B$ brings also closer the evanescent waves which are paired in the state, raising its energy gradually until it gets a sensible nonvanishing value. Thus, for very large magnetic fields (with $B \gtrsim 60$ T), a gap opens up in the zero Landau level and the massive degeneracy is lost.

It is remarkable that the huge degeneracy of the zero Landau level affords a quantization of the Hall conductivity along the opposite faces of the slab. This can be shown by observing that the current density along the $x$ direction, $j_x = -2m_1 (k_x + Bz) \sigma_z $, is given by the derivative $\partial H_{\parallel }/\partial k_x$. Thus, the current of each state across the whole section of the slab (for a finite lateral dimension $\Delta y$) is given in terms of the dispersion $\varepsilon (k_x)$ by 
\begin{equation}
\frac{1}{W} \frac{1}{\Delta y} \int dy dz \: \chi^\dagger j_x \chi = \frac{\partial \varepsilon }{\partial k_x}
\end{equation}
Considering the situation in which the Fermi level is right above the zero Landau level, we obtain the intensity $I_x$ along the $x$ direction by integrating over all the filled states in the bands dispersing from zero energy. We get (reinstating at this point $\hbar $ in the equations)
\begin{equation}
I_x = \frac{e}{\hbar } \int_{\mbox{\rm {\footnotesize filled  states}}}  \frac{dk_x}{2\pi } \frac{\partial \varepsilon }{\partial k_x}
\label{ix}
\end{equation}\\
The integral at the right-hand side of (\ref{ix}) is nothing but the difference between the respective chemical potentials $\varepsilon_+, \varepsilon_-$ at the two opposite faces of the slab, times the degeneracy $N$ of the zero Landau level. Therefore, we have $I_x = N (e/h) (\varepsilon_+ - \varepsilon_-)$ and the transverse Hall conductivity $\sigma_{xz} = N (e^2/h)$, making clear that this is built entirely from the currents carried by the quasi-2D states at opposite surfaces of the slab.

{\em Conclusions.---}
We have shown that, under a strong magnetic field, a 3D nodal-line semimetal enters a topological insulating phase in which the low-energy states in the bulk do not contribute to the conductivity and all the electronic transport takes place at the surface of the material. 
%A nodal line semimetal in the presence of a magnetic field shows very interesting physics depending on the geometry of the experimental configuration. In particular, in the case of a magnetic field following the axis of symmetry of the nodal line structure, the drumhead states evolve into Hall states in the surface. On the contrary, if the magnetic field is in the same plane as the nodal ring, there appears a 3D quantum Hall effect in which Landau states are extended on layers of the bulk. 

In the 3D quantum Hall effect we have unveiled, a key feature is the presence of a zero Landau level with huge degeneracy, implying a very high but quantized Hall conductivity $\sigma_{ij} = N (e^2/h)$ with large values of the Chern number $N$. When the magnetic field is parallel to the nodal ring, $N$ increases linearly with the width of the surface, typically at a rate of a new channel per each few nanometers along the section of the surface. Thus, it should be possible to create a macroscopic surface current with $N \gtrsim 100$ in samples of micron scale.

%However, the realistic values of $N$ we can achieve vary very much in both cases. For the surface Hall efect for the magnetic field in the parallel configuration one can estimate $N$ for the family of CaP$_3$ compounds taken parameter values from DFT calculations published in Ref. \cite{Xu16}. For example, for the CaP$_3$ case at $B=5 T$ we estimate values of $N \approx 4$. This number depends greatly on the ratio $m_0/m_1$ and can vary very much in different materials. For the 3D Hall effect the value of $N$ depends not only on the magnetic field and the material parameters but also increases with the width of the sample and can potentially reach much higher values.  

A most remarkable effect of the predicted topological insulating phase is that the orientation of the surface currents is dictated exclusively by the direction of the magnetic field. In this setup, currents can only flow unidirectionally along opposite surfaces of the 3D material, as a reflection of surface states which are immune to backscattering, and showing at work the topological protection of the 3D Landau states from the nodal-line semimetal.

%The Hall conductivity is proportional to the degeneracy of the zero mode state. It is possible to estimate this Chern number $N$ depending on the strength of the magnetic field for the family of CaP$_3$ compounds taken parameter values from DFT calculations published in Ref. \cite{Xu16} for both the parallel (surface Hall effect) and perpendicular (3D Hall effect). 
% rounding  and ignoring anisotropies m0=0.15 eV, m1=3 eV·nm^2, v=1 eV·nm
%For example, for a CaP$_3$ wire with section $40 \times 40$ nm and $B=30 T$ $N$ would be:   

\acknowledgments

We acknowledge financial support through Spanish grants MINECO/FEDER No. FIS2015-63770-P and No. FIS2014-57432-P.


\begin{thebibliography}{xx}
\bibitem{Liu14}
Z. K. Liu, B. Zhou, Y. Zhang, Z. J. Wang, H. M. Weng, D. Prabhakaran, S.-K. Mo, Z. X. Shen, Z. Fang, X. Dai, Z. Hussain and Y. L. Chen, Discovery of a Three-Dimensional Topological Dirac Semimetal, Na$_3$Bi, Science {\bf 343}, 864 (2014).

\bibitem{Neupane14}
M. Neupane, S. Xu, R. Sankar, N. Alidoust, G. Bian, C. Liu, I. Belopolski, T.-R. Chang, H.-T. Jeng, H. Lin, A. Bansil, F. Chou and M. Z. Hasan, Observation of a three-dimensional topological Dirac semimetal phase in high-mobility Cd$_3$As$_2$, Nature Commun. {\bf 5}, 3786 (2014).

\bibitem{Borisenko14}
S. Borisenko, Q. Gibson, D. Evtushinsky, V. Zabolotnyy, B. B\"uchner and R. J. Cava, Experimental realization of a three-dimensional Dirac semimetal, Phys. Rev. Lett. {\bf 113}, 027603 (2014).

\bibitem{Xu15}
S.-Y. Xu, I. Belopolski, N. Alidoust, M. Neupane, G. Bian, C. Zhang, R. Sankar, G. Chang, Z. Yuan, C.-C. Lee, S.-M. Huang, H. Zheng, J. Ma, D. S. Sanchez, B. Wang, A. Bansil, F. Chou, P. P. Shibayev, H. Lin, S. Jia, and M. Z. Hasan, Discovery of a Weyl fermion semimetal and topological Fermi arcs,
Science {\bf 349}, 613 (2015).

\bibitem{Burkov16} A.A. Burkov, Topological Semimetals, Nat. Materials {\bf 15}, 1145 (2016).



\bibitem{Fang16} C. Fang, H. Weng, X. Dai, Z. Fang, Topological nodal line semimetals, Chin. Phys. B {\bf 25}, 117106 (2016).

\bibitem{Bian16} G. Bian {\em et al.}, Topological nodal-line fermions in spin-orbit metal PBTaSe$_2$, Nature Commun. {\bf 7}, 10556  (2016).

%Guang Bian, Tay-Rong Chang, Raman Sankar, Su-Yang Xu, Hao Zheng, Titus Neupert, Ching-Kai Chiu, Shin-Ming Huang, Guoqing Chang, Ilya Belopolski, Daniel S. Sanchez, Madhab Neupane, Nasser Alidoust, Chang Liu, BaoKai Wang, Chi-Cheng Lee, Horng-Tay Jeng, Chenglong Zhang, Zhujun Yuan, Shuang Jia, Arun Bansil, Fangcheng Chou, Hsin Lin & M. Zahid Hasan

\bibitem{Wu16} Y. Wu, L.-L. Wang, E. Mun, D. D. Johnson, D. Mou, L. Huang, Y. Lee, S. L. Bud'ko, P. C. Canfield, A. Kaminski, Dirac node arcs in PtSn$_4$, Nature Phys. {\bf 12}, 667 (2016). 

\bibitem{Schoop16} L. M. Schoop, M. N. Ali, C. Straßer, V. Duppel, S. S. P. Parkin, B. V. Lotsch, and C. R.
Ast, Dirac Cone Protected by Non-Symmorphic Symmetry and 3D Dirac Line Node in ZrSiS, Nature Commun. {\bf 7}, 11696 (2015). 

\bibitem{Jia16} S. Jia, S.-Y. Zu, M. Z. Hasan, Weyl semimetals, Fermi arcs and chiral anomaly, Nature Mat. {\bf 15}, 1140 (2016).

\bibitem{Klitzing80} K. von Klitzing, G. Dorda, M. Pepper, New method for high-accuracy determination of the fine-structure constant based on quantized Hall resistance, Phys. Rev. Lett. {\bf 45}, 494 (1980).

\bibitem{Thouless82} D. Thouless, M. Kohmoto, M. Nightingale, M. den Nijs, Phys. Rev. Lett. {\bf 49}, 405 (1982).

%\bibitem{Novoselov07} K.S. Novoselov, Z. Jiang, Y. Zhang, S.V. Morozov, H.L. Stormer, U. Zeitler, J.C. Maan, G.S. Boebinger, Room-Temperature Quantum Hall Effect in Graphene, Science 315, 1379 (2007). 

\bibitem{Yang11} K.-Y. Yang, Y.-M. Lu, Y. Ran, Quantum Hall effects in a Weyl semimetal: Possible application in pyrochlore iridates, Phys. Rev. B {\bf 84}, 075129 (2011).

\bibitem{Cao15} J. Cao, S. Liang, C. Zhang, Y. Liu, J. Huang, Z. Jin, Z.-G. Chen, Z. Wang, Q. Wang, J. Zhao, S. Li, X. Dai, J. Zou, Z. Xia, L. Li, F. Xiu, Landau level splitting in Cd$_3$As$_2$ under high magnetic fields, Nature Commun. {\bf 6}, 7779 (2015).

\bibitem{Potter14} A.C. Potter, I. Kimchi, A. Vishwanath, Surface to bulk Fermi arcs via Weyl nodes as topological defects, Nature Comm. {\bf 5}, 5161 (2014). % Is this the right reference?

\bibitem{Rhim15} J.-W. Rhim, Y. B. Kim, Landau level quantization and almost flat modes in three dimensional semimetals with nodal ring spectra, Phys. Rev. B {\bf 92}, 045126 (2015).

\bibitem{Berry04} M.V. Berry, Physics of non-Hermitian degeneracies, Czech. J. Phys. {\bf 54}, 1039 (2004).
 
\bibitem{Heiss12} W.D. Heiss, The physics of exceptional points, J. Phys. A: Math. Theor. {\bf 45}, 444016 (2012).

\bibitem{Gonzalez17} J. Gonz\'alez, R.A. Molina, Topological protection from exceptional points in Weyl and nodal-line semimetals, Phys. Rev. B {\bf 96}, 045437 (2017).

\bibitem{Gonzalez16} J. Gonz\'alez, R.A. Molina, Macroscopic degeneracy of zero-mode rotating surface states in 3D Dirac and Weyl semimetals under radiation, Phys. Rev. Lett. {\bf 116}, 156803 (2016).

\bibitem{Xu16} Z. Xu, R. Yu, Z. Fang, X. Dai, H. Weng, Topological nodal line semimetals in CaP$_3$ family of materials, Phys. Rev. B {\bf 95}, 045136 (2017).

\end{thebibliography}
\end{document}